# A Cu$_3$BHT-Graphene van der Waals Heterostructure with Strong Interlayer Coupling

Zhiyong Wang[1,2], Shuai Fu[2,3], Wenjie Zhang[1], Baokun Liang[4], Tsai-Jung Liu[2], Mike Hambsch[5], Jonas F. Pöhls[6], Yufeng Wu[1], Jianjun Zhang[2], Tianshu Lan[1], Xiaodong Li[1,2], Haoyuan Qi[4], Miroslav Polozij[2,9], Stefan C. B. Mannsfeld[5], Ute Kaiser[4], Mischa Bonn[3], R. Thomas Weitz[6], Thomas Heine[2,9,10], Stuart S. P. Parkin[1]*, Hai I. Wang[3,7]*, Renhao Dong[2,8]*, Xinliang Feng[1,2]*

[1]Max Planck Institute for Microstructure Physics, Halle (Saale), Germany.
[2]Center for Advancing Electronics Dresden (cfaed) and Faculty of Chemistry and Food Chemistry, Technische Universität Dresden, Dresden, Germany.
[3]Max Planck Institute for Polymer Research, Mainz, Germany.
[4]Central Facility for Electron Microscopy, Electron Microscopy of Materials Science, Ulm University, Ulm, Germany.
[5]Center for Advancing Electronics Dresden, Technische Universität Dresden, Dresden, Germany.
[6]First Institute of Physics, Georg August University of Göttingen, Göttingen, Germany.
[7]Nanophotonics, Debye Institute for Nanomaterials Science, Utrecht University, Utrecht, the Netherlands.
[8]Key Laboratory of Colloid and Interface Chemistry of the Ministry of Education, School of Chemistry and Chemical Engineering, Shandong University, Jinan, China.
[9]Helmholtz-Zentrum Dresden-Rossendorf, Institute of Resource Ecology, Leipzig, Germany.
[10]Department of Chemistry, Yonsei University, Seoul, Republic of Korea.
*E-mail: stuart.parkin@mpi-halle.mpg.de; h.wang05@uu.nl; renhaodong@sdu.edu.cn; xinliang.feng@tu-dresden.de

**Two-dimensional van der Waals heterostructures (2D vdWhs) are of significant interest due to their intriguing physical properties that are critically defined by the constituent monolayers and their interlayer coupling. However, typical inorganic 2D vdWhs fall into the weakly coupled region, limiting efficient interfacial charge flow—crucial for developing high-performance quantum optoelectronics. Here, we demonstrate strong interlayer coupling in Cu$_3$BHT (BHT = benzenehexathiol)-graphene vdWhs, an organic-inorganic bilayer characterized by prominent interlayer charge transfer. Monolayer Cu$_3$BHT with a Kagome lattice is synthesized on the water surface and then coupled with graphene to produce a cm$^2$-scale 2D vdWh. Spectroscopic and electrical studies, along with theoretical calculations, show significant hole transfer from monolayer Cu$_3$BHT to graphene upon contact, being characteristic fingerprints for strong interlayer coupling. This study unveils the great potential of integrating highly π-conjugated 2D coordination polymers (2DCPs) into 2D vdWhs to explore intriguing physical phenomena.**

Two-dimensional van der Waals heterostructures (2D vdWhs) constitute a new class of artificial materials constructed by vertically stacking atomically thin 2D crystals, including graphene, transition metal dichalcogenides (TMDs), and hexagonal boron nitride (h-BN) which are held together by vdW interactions[1,2]. The lack of direct chemical bonding in vdWhs affords complete freedom in the selection of 2D materials, unrestricted by constraints of crystal symmetry, structure, and lattice matching[3]. Thanks to their atomically thin nature, the properties of vdWhs can be tailored by not only individual constituents but also interlayer coupling, giving rise to new quantum states and phenomena[4]. These unique characteristics have established vdWhs as a versatile platform for a wide spectrum of applications, including tunneling transistors[5], photodetectors,[6] and light-emitting devices[7] with unprecedented functionalities. While vdWhs hold great promise for developing new materials and device concepts, the fabrication of high-quality monolayer 2D crystals is limited by the exfoliation of existing bulk-layered crystals, which restricts the scope and progress of vdWhs research. In addition, further exploration of 2D vdWhs is still impeded by the usually weak interlayer coupling and the lack of control in tuning the interlayer interactions[8-10]. Hence, the discovery of novel vdWhs that enable manipulation of band alignment and interlayer electronic coupling is a topic of great research interest.

Here, we exploit the synergistic effect of highly delocalized electronic orbitals in $Cu_3BHT$ and massless Dirac Fermions in graphene to target strongly coupled organic-inorganic bilayers. We show that vertically stacking monolayer $Cu_3BHT$ and graphene through a wet transfer technique enables the creation of $cm^2$-scale $Cu_3BHT$-graphene vdWhs with intimate and uniform contacts. Raman and terahertz time-domain spectroscopy (THz-TDS) studies, combined with electrical measurements and theoretical modeling, demonstrate significant hole transfer from monolayer $Cu_3BHT$ to graphene upon contact, leading to a downshift of the Fermi level ($E_F$) in graphene arising from the strong interlayer electronic coupling. These findings open up new possibilities for exploring novel 2D vdWhs with manipulable orbital distribution, interfacial band alignment, and interlayer coupling, a cornerstone for developing and envisioning novel quantum optoelectronics.

**Results and discussion**

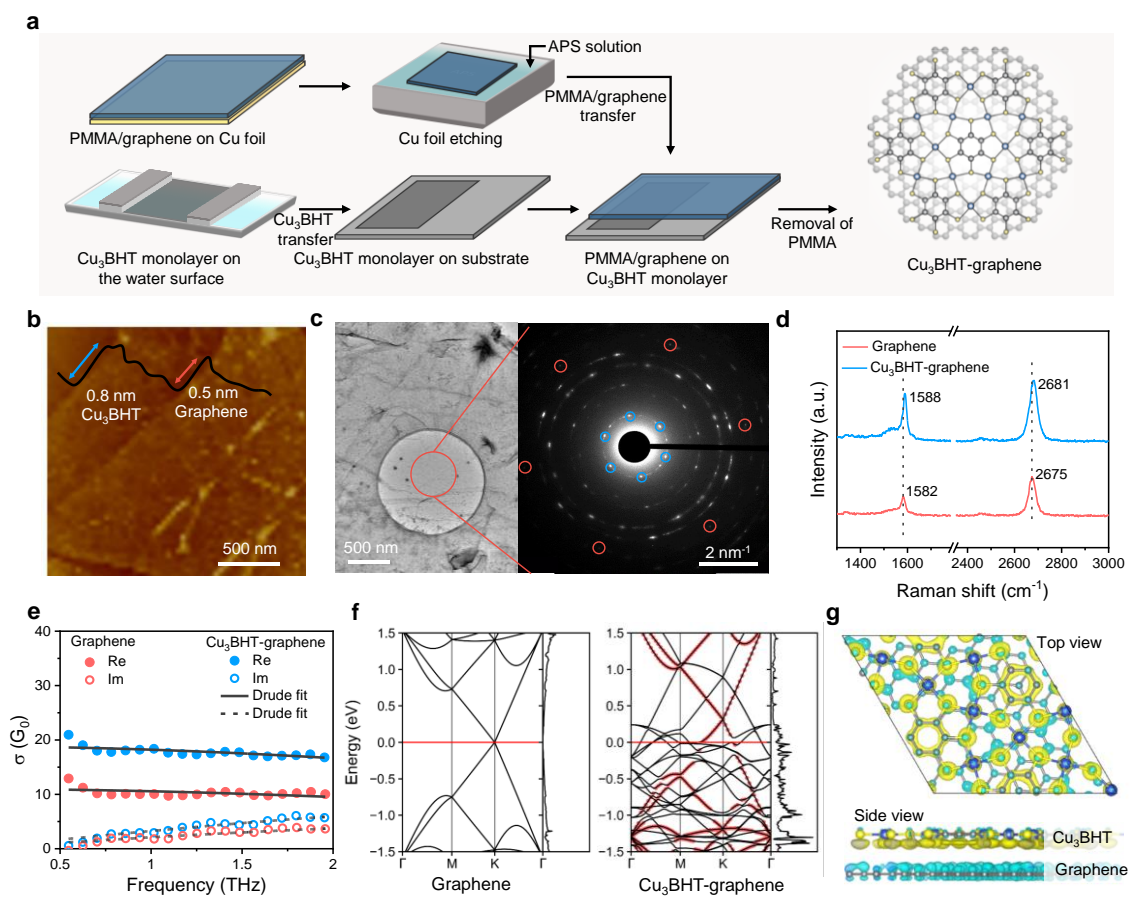

**Figure 1| Morphological and interlayer coupling characterization of $Cu_3BHT$-graphene**

**vdWh. a**, Schematic of the preparation of $Cu_3BHT$-graphene vdWh by a wet transfer technique. **b**, AFM image and height profile of monolayer $Cu_3BHT$ (blue line) and graphene (yellow line) in the vdWh. **c**, TEM image and corresponding SAED pattern of $Cu_3BHT$-graphene vdWh. **d**, Raman spectra of graphene before and after deposition of monolayer $Cu_3BHT$. In the vdWh, the well-preserved symmetry of the G- and 2D-modes indicates that graphene maintains its structural integrity after contact with $Cu_3BHT$. **e**, Frequency-resolved complex sheet conductivity $\sigma_s(w)$ of bare graphene (red) and $Cu_3BHT$-graphene vdWh (blue) in units of quantum conductance $G_0 = 2e^2/h$. The solid and dashed lines correspond to the Drude fits of the real and imaginary parts of $\sigma_s(w)$, respectively. **f**, Band structures of graphene and graphene-$Cu_3BHT$ vdWh with 3 °twist angle and their corresponding densities of states (DOS). Red dots in graphene-$Cu_3BHT$ vdWh band structure indicate bands originating predominantly from the graphene layer. **g**, The CDD of $Cu_3BHT$-graphene vdWh. Charge density isosurfaces value is set to 0.0003 e·$Bohr^{-3}$ with yellow and blue showing increase and decrease of electron density, respectively.

We fabricated the vdWh by combining monolayer $Cu_3BHT$ synthesized on the water surface with large-area chemical vapor deposited (CVD) graphene using a wet transfer technique (Fig. 1a). The resulting vdWh was then washed and annealed at 300 °C under $H_2$-Ar atmosphere to remove PMMA and trapped interfacial contaminants, and to enhance the interaction between $Cu_3BHT$ and graphene. Fig. 1b shows the morphology of $Cu_3BHT$-graphene vdWh, from which the monolayer nature of both $Cu_3BHT$ and graphene layers was confirmed. We further characterized the structures of $Cu_3BHT$-graphene vdWh using SAED and TEM (Fig. 1c, left). The SAED pattern from the overlapping $Cu_3BHT$-graphene region shows two sets of diffraction spots (Fig. 1c, right), with the inner and outer sets corresponding to $Cu_3BHT$ and graphene, respectively. Note that no clear Moiré superlattice structure was observed using AC-HRTEM imaging, mainly due to the limited domain size (tens of nm) of the polycrystalline $Cu_3BHT$[11-14].

We conducted Raman and THz measurements to study the electron re-equilibrium upon vdWh formation. Raman spectra of graphene, taken before and after deposition of monolayer $Cu_3BHT$, were averaged from Raman mapping images and presented in Fig. 1d. Pristine graphene exhibits G- and 2D-modes at 1,582 $cm^{-1}$ and ~2,675 $cm^{-1}$, respectively, indicative of high-quality monolayer graphene with negligible strain and doping effects[15,16]. The deposition of monolayer $Cu_3BHT$ resulted in a blue-shift (~6 $cm^{-1}$) of both G- (~1,588 $cm^{-1}$) and 2D (2,681 $cm^{-1}$)-modes, indicating an increased hole concentration in graphene[15]. By deconvoluting the contributions of strain and doping using the correlation between G-mode and 2D-mode[16], we find that monolayer $Cu_3BHT$ coverage on graphene introduces negligible strain but substantial hole density modulation, whereby the $E_F$ in graphene shifts downward by ~0.29 eV in vdWh (equivalent to a hole concentration of ~5×$10^{12}$ $cm^{-2}$).

To gain deeper insight into the ground-state charge transfer (CT) effect upon vdWh formation, we employ time-resolved THz spectroscopy (TRTS) to investigate the frequency-resolved complex sheet conductivity $\sigma(\omega)$ of graphene with and without monolayer $Cu_3BHT$ interactions. Given the negligible intrinsic conductivity of monolayer $Cu_3BHT$, the observed THz absorption (and thus conductivity) in bare graphene and $Cu_3BHT$-graphene vdWh can be attributed to the presence of conductive charge carriers in the graphene layer. As shown in Fig. 1e, $\sigma(\omega)$ of bare graphene and $Cu_3BHT$-graphene vdWh can be well described by the Drude model (equation 1)[17,18].

$$\sigma(\omega) = \frac{D}{\pi\left(\frac{1}{\tau}-i\omega\right)} \quad \text{...............................................(1)}$$

where $\omega$ is the angular frequency, $D$ is the Drude weight related to the carrier density $N$ and $E_F$, and $\tau$ is the charge scattering time. The fits infer $\tau$ and $E_F$ to be 31 ±2 fs and ~0.23 eV for bare graphene, and 29 ±1 fs and ~0.43 eV for monolayer $Cu_3BHT$-covered graphene. These findings indicate a downshift of $E_F$ in graphene by ~0.20 eV upon vdWh formation, corroborating the Raman results.

To further explore the electrical characteristics evolution upon vdWh formation, we calculated their band structures (Fig. 1f) and electrostatic potentials using DFT. The calculated $E_F$ of the vdWh is positioned 0.315 eV below the graphene Dirac cone, signifying a prominent hole accumulation in graphene. Furthermore, we tested three different twist angles, all of which yielded consistent qualitative results, suggesting that the observed effect is independent of interlayer twist and, consequently, the domain position within the polycrystalline material. The charge density difference (CDD) isosurfaces highlight a strong interaction between monolayer $Cu_3BHT$ and graphene (Fig. 1g), leading to electron accumulation in $Cu_3BHT$ and hole accumulation in graphene. These results align with our Raman and TRTS observations and identify that the strong interlayer coupling contributes to significant hole transfer from monolayer $Cu_3BHT$ to graphene upon contact.

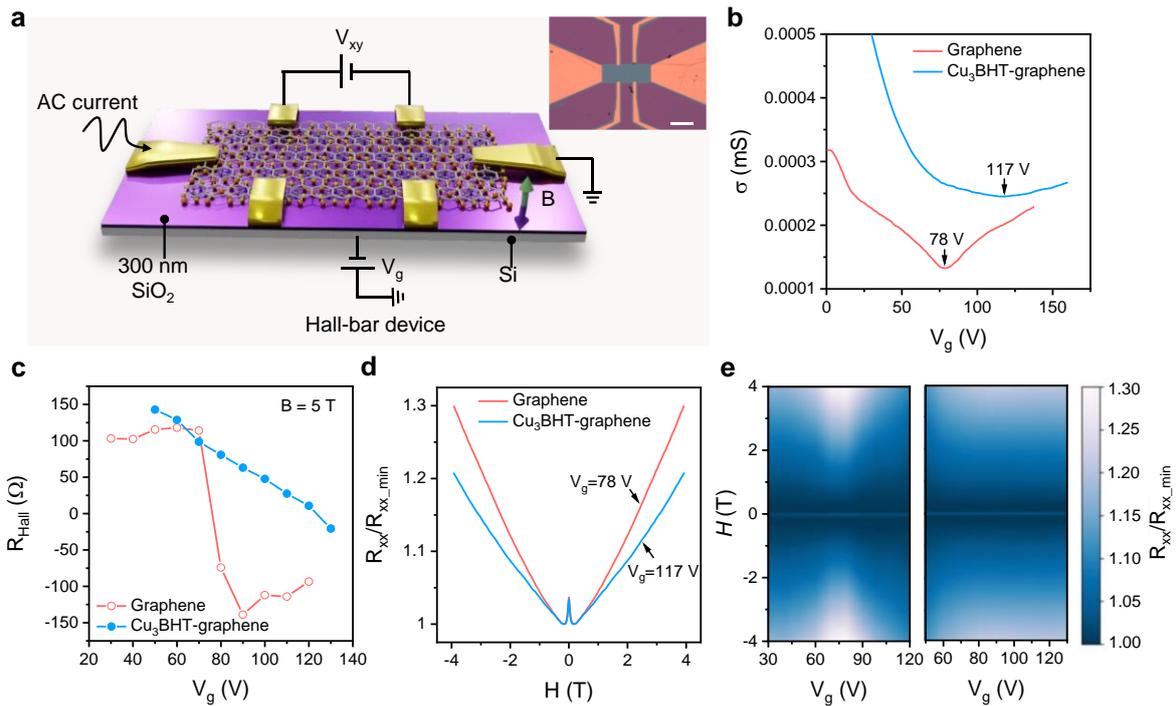

**Figure 2| Electrical study of CT in $Cu_3BHT$-graphene vdWh. a**, Schematic illustration of the gated-Hall bar device. Inset: Optical image of the gated-Hall bar device. Scale bar, 50 μm. **b**, Gate-voltage-dependent longitudinal conductance of pristine graphene and $Cu_3BHT$-graphene vdWh at 2 K, respectively. As the gate voltage increases, the $E_F$ in graphene passes through the Dirac point, leading to a transition in the type of graphene from hole accumulation to electron accumulation. The conductance in graphene initially decreases as the $E_F$ approaches the Dirac point, followed by an increase when the $E_F$ departs from the Dirac point. **c**, Gate-voltage-dependent Hall resistance of pristine graphene and $Cu_3BHT$-graphene vdWh under 5 T magnetic field, respectively. **d**, Normalized longitudinal resistances $R_{xx}/R_{xx\_min}$ of

graphene and Cu$_3$BHT-graphene vdWh versus field, with applied gate voltage 78 V and 117 V, respectively. A broadened valley manifests that the incorporation of Cu$_3$BHT indeed impacts the band structure of graphene. **e**, Mapping of pristine graphene and Cu$_3$BHT-graphene vdWh's normalized longitudinal resistances $R_{xx}/R_{xx\_min}$, measured with varied magnetic fields and different gate voltages.

We further fabricated gated-Hall bar devices to investigate the electronic properties of Cu$_3$BHT-graphene vdWh under external electric and magnetic fields (Fig. 2a). As shown in Fig. 2b, the conductance-gate voltage curve of graphene exhibits a distinct minimum at a gate voltage of ~78 V, which corresponds to the Dirac point. In contrast, the Cu$_3$BHT-graphene vdWh displays a broader valley in the curve, positioned around a gate voltage of ~117 V. This significant shift in the Dirac point suggests the transfer of holes from Cu$_3$BHT to the graphene layer[19], requiring a higher gate voltage to raise the $E_F$ to the Dirac point. Fig. 2c shows the gate voltage-dependent Hall coefficients, which exhibit reversals of signs at the Dirac points due to the change from hole accumulation to electron accumulation. In pristine graphene, either holes or electrons dominate the Hall signals, leading to a sharp reversal. However, in Cu$_3$BHT-graphene vdWh, both holes and electrons contribute to the Hall signal, with their respective contributions varying as a function of gate voltage, resulting in a gradual reversal. We further investigate the magnetic field dependence of normalized longitudinal resistance $R_{xx}/R_{xx\_min}$, as presented in Fig. 2d. These curves are interpolated by data points measured at the neighborhoods of their Dirac points. Close to zero field, a small peak is observed, which could be related to weak localization[20]. The remaining curves exhibit positive magneto-resistance, consistent with the Drude model. The full mapping of gate voltage-dependent magneto-resistance of graphene and Cu$_3$BHT-graphene vdWh are depicted in Fig. 2e. In graphene, a 130% positive magneto-resistance is observed around the Dirac point, which diminishes significantly as the system moves away from the Dirac point, aligning with previous reports[21]. However, in Cu$_3$BHT-graphene vdWh, the magneto-resistance remains largely unchanged over a wide range of gate voltages. Collectively, these observations further identify the strong interlayer coupling at the Cu$_3$BHT-graphene interface.

**Discussion**

In conclusion, we have demonstrated strong interlayer coupling in a novel Cu$_3$BHT-graphene vdWh. Vertically stacking monolayer Cu$_3$BHT and graphene through a wet transfer technique allows facile assembly of Cu$_3$BHT-graphene vdWhs on the cm$^2$ scale with intimate and uniform contacts. As revealed by optical and electrical studies as well as theoretical calculations, the strong interlayer coupling leads to a substantial amount of holes flowing from monolayer Cu$_3$BHT to graphene upon contact. Our findings lay the foundation for the creation of innovative vdWhs with modulated band alignment and interlayer orbital coupling by integrating monolayer 2DCPs with inorganic 2D materials (e.g., graphene, TMDs, and *h*-BN), and will propel the exploitation of intriguing physical phenomena and specific optoelectronic devices.

**Acknowledgments**

This work is financially supported by EU Graphene Flagship (GrapheneCore3, No. 881603), ERC starting grant (FC2DMOF, No. 852909), ERC Consolidator Grant (T2DCP), DFG project (2D polyanilines, No. 426572620), H2020-FETOPEN (PROGENY, 899205), CRC 1415 (Chemistry of Synthetic Two-Dimensional Materials, No. 417590517), SPP 2244 (2DMP), GRK2861 (No. 491865171), EMPIR-20FUN03-COMET, as well as the German Science

Council and Center of Advancing Electronics Dresden (cfaed). R.D. thanks Taishan Scholars Program of Shandong Province (tsqn201909047) and National Natural Science Foundation of China (22272092). The authors acknowledge cfaed and Dresden Center for Nanoanalysis (DCN) at TUD and Dr. Petr Formanek, Prof. Andreas Fery, Anna Maria Dominic and Prof. Inez M. Weidinger for the use of TEM facility at IPF, as well as the Raman measurement. We acknowledge SOLEIL for provision of synchrotron radiation facilities and we would like to thank Dr. Arnaud Hemmerle for assistance in using beamline SIRIUS. T.L., M.P. and T.H. thank ZIH Dresden for the use of computational resources.